%%%%%%%%%%%%%%%%%%%%%%%%%%%%%%%%%%%%%%%%%%%%%%%%%%%%%%%%%%%%%%%%%%%%%%%%%%%%%%%
%%%  TeX harvmac
%%%  D6 branes and M-theory geometrical transitions from gauged supergravity
%%%  Jose D. Edelstein and Carlos Nunez
%%%%%%%%%%%%%%%%%%%%%%%%%%%%%%%%%%%%%%%%%%%%%%%%%%%%%%%%%%%%%%%%%%%%%%%%%%%%%%%
\input harvmac

\input psfig
\newcount\figno
\figno=0
\def\fig#1#2#3{
\par\begingroup\parindent=0pt\leftskip=1cm\rightskip=1cm\parindent=0pt
\global\advance\figno by 1
\midinsert
\epsfxsize=#3
\centerline{\epsfbox{#2}}
\vskip 12pt
{\bf Fig. \the\figno:} #1\par
\endinsert\endgroup\par
}
\def\figlabel#1{\xdef#1{\the\figno}}
\def\encadremath#1{\vbox{\hrule\hbox{\vrule\kern8pt\vbox{\kern8pt
\hbox{$\displaystyle #1$}\kern8pt}
\kern8pt\vrule}\hrule}}

\overfullrule=0pt

%% MACROS

% Something to deal with sub-sub-sections

\def\unlockat{\catcode`\@=11}
\def\lockat{\catcode`\@=12}

\unlockat
% Something to deal with sub-sub-sections

\def\newsec#1{\global\advance\secno by1\message{(\the\secno. #1)}
\global\subsecno=0\global\subsubsecno=0\eqnres@t\noindent
{\bf\the\secno. #1}
\writetoca{{\secsym} {#1}}\par\nobreak\medskip\nobreak}
\global\newcount\subsecno \global\subsecno=0
\def\subsec#1{\global\advance\subsecno
by1\message{(\secsym\the\subsecno. #1)}
\ifnum\lastpenalty>9000\else\bigbreak\fi\global\subsubsecno=0
\noindent{\it\secsym\the\subsecno. #1}
\writetoca{\string\quad {\secsym\the\subsecno.} {#1}}
\par\nobreak\medskip\nobreak}
\global\newcount\subsubsecno \global\subsubsecno=0
\def\subsubsec#1{\global\advance\subsubsecno by1
\message{(\secsym\the\subsecno.\the\subsubsecno. #1)}
\ifnum\lastpenalty>9000\else\bigbreak\fi
\noindent\quad{\secsym\the\subsecno.\the\subsubsecno.}{#1}
\writetoca{\string\qquad{\secsym\the\subsecno.\the\subsubsecno.}{#1}}
\par\nobreak\medskip\nobreak}

\def\subsubseclab#1{\DefWarn#1\xdef
#1{\noexpand\hyperref{}{subsubsection}%
{\secsym\the\subsecno.\the\subsubsecno}%
{\secsym\the\subsecno.\the\subsubsecno}}%
\writedef{#1\leftbracket#1}\wrlabeL{#1=#1}}% Macros for boxes
\lockat

\def\IL{\relax{\rm I\kern-.18em L}}
\def\IH{\relax{\rm I\kern-.18em H}}
\def\IR{\relax{\rm I\kern-.18em R}}
\def\IC{\relax\hbox{$\inbar\kern-.3em{\rm C}$}}
\def\IT{\relax\hbox{$\inbar\kern-.3em{\rm T}$}}
\def\IZ{\relax\ifmmode\mathchoice
{\hbox{\cmss Z\kern-.4em Z}}{\hbox{\cmss Z\kern-.4em Z}}
{\lower.9pt\hbox{\cmsss Z\kern-.4em Z}}
{\lower1.2pt\hbox{\cmsss Z\kern-.4em Z}}\else{\cmss Z\kern-.4em
Z}\fi}

\def\CN {{\cal N}}

\def\CD {{\cal D}}

\def\CL {{\cal L}}

\def\CO {{\cal O}}

\def\CS {{\cal S}}

%% MORE MACROS

\def\CN {{\cal N}}

\def\CO {{\cal O}}

\def\CS {{\cal S }}

\font\manual=manfnt \def\dbend{\lower3.5pt\hbox{\manual\char127}}

\def\IZ{\relax\ifmmode\mathchoice
{\hbox{\cmss Z\kern-.4em Z}}{\hbox{\cmss Z\kern-.4em Z}}
{\lower.9pt\hbox{\cmsss Z\kern-.4em Z}}
{\lower1.2pt\hbox{\cmsss Z\kern-.4em Z}}\else{\cmss Z\kern-.4em
Z}\fi}
\def\half {{1\over 2}}

\def\p{\partial}

\def\CN {{\cal N}}

\def\CO {{\cal O}}

\def\CS {{\cal S }}

% more macros, alphabetically

\def\IZ{\relax\ifmmode\mathchoice
{\hbox{\cmss Z\kern-.4em Z}}{\hbox{\cmss Z\kern-.4em Z}}
{\lower.9pt\hbox{\cmsss Z\kern-.4em Z}}
{\lower1.2pt\hbox{\cmsss Z\kern-.4em Z}}\else{\cmss Z\kern-.4em
Z}\fi}
\def\IB{\relax{\rm I\kern-.18em B}}
\def\IC{{\relax\hbox{$\inbar\kern-.3em{\rm C}$}}}
\def\ID{\relax{\rm I\kern-.18em D}}
\def\IE{\relax{\rm I\kern-.18em E}}
\def\IF{\relax{\rm I\kern-.18em F}}
\def\IG{\relax\hbox{$\inbar\kern-.3em{\rm G}$}}
\def\IGa{\relax\hbox{${\rm I}\kern-.18em\Gamma$}}
\def\IH{\relax{\rm I\kern-.18em H}}
\def\II{\relax{\rm I\kern-.18em I}}
\def\IK{\relax{\rm I\kern-.18em K}}
\def\IP{\relax{\rm I\kern-.18em P}}
\def\IQ{\relax\hbox{$\inbar\kern-.3em{\rm Q}$}}

\def\inbar{\,\vrule height1.5ex width.4pt depth0pt}

\def\p{\partial}

\font\cmss=cmss10 \font\cmsss=cmss10 at 7pt
\def\IR{\relax{\rm I\kern-.18em R}}

% Macros for boxes

\def\boxit#1{\vbox{\hrule\hbox{\vrule\kern8pt
\vbox{\hbox{\kern8pt}\hbox{\vbox{#1}}\hbox{\kern8pt}}
\kern8pt\vrule}\hrule}}
\def\mathboxit#1{\vbox{\hrule\hbox{\vrule\kern8pt\vbox{\kern8pt
\hbox{$\displaystyle #1$}\kern8pt}\kern8pt\vrule}\hrule}}

%% ANOTHER SET OF MACROS

\def\inbar{\,\vrule height1.5ex width.4pt depth0pt}

\def\p{\partial}

\font\cmss=cmss10 \font\cmsss=cmss10 at 7pt
\def\IR{\relax{\rm I\kern-.18em R}}

\def\a{{\alpha}}
\def\b{{\beta}}
\def\d{{\delta}}
\def\e{{\epsilon}}
\def\g{{\gamma}}
\def\kap{{\kappa}}
\def\la{{\lambda}}
\def\o{{\omega}}
\def\ph{{\phi}}
\def\s{{\sigma}}
\def\t{{\theta}}
\def\G{{\Gamma}}
\def\O{{\Omega}}
\def\tw{{\tilde w}}
\def\tS{{\tilde S}}
\def\tL{{\tilde L}}

\def\ur{{\underline{r}}}
\def\ut{{\underline{\theta}}}
\def\up{{\underline{\phi}}}
\def\ups{{\underline{\psi}}}

%% new macros

%% END MACROS

\lref\amv{M.~Atiyah, J.~Maldacena and C.~Vafa, ``An M-theory flop as a
large N duality'', hep-th/0011256.}
%%CITATION = HEP-TH 0011256;%%

\lref\ss{A.~Salam and E.~Sezgin, ``d=8 Supergravity'', Nucl.\ Phys.\ B {\bf
258} (1985) 284.}
%%CITATION = NUPHA,B258,284;%%

\lref\scs{J.~Scherk and J.~H.~Schwarz, ``How To Get Masses From Extra
Dimensions'', Nucl.\ Phys.\ B {\bf 153} (1979) 61.}
%%CITATION = NUPHA,B153,61;%%

\lref\agk{B.~S.~Acharya, J.~P.~Gauntlett and N.~Kim, ``Fivebranes wrapped
on associative three-cycles'', hep-th/0011190. ~H.~Nieder and Y.~Oz,
``Supergravity and D-branes wrapping special Lagrangian cycles'',
JHEP{\bf 0103} (2001) 008 [hep-th/0011288]. ~J.~P.~Gauntlett, N.~Kim and
D.~Waldram, ``M-fivebranes wrapped on supersymmetric cycles'',
hep-th/0012195. ~C.~N\'u\~nez, I.~Y.~Park, M.~Schvellinger and T.~A.~Tran,
``Supergravity duals of gauge theories from F(4) gauged supergravity in
six dimensions'', hep-th/0103080.}
%%CITATION = HEP-TH 0011190;%%
%%CITATION = HEP-TH 0011288;%%
%%CITATION = HEP-TH 0012195;%%
%%CITATION = HEP-TH 0103080;%%

\lref\mnu{J.~M.~Maldacena and C.~N\'u\~nez, ``Supergravity description of
field theories on curved manifolds and a no go theorem'', hep-th/0007018.
~``Towards the large N limit of pure $\CN = 1$ super Yang Mills'', Phys.\
Rev.\ Lett.\ {\bf 86} (2001) 588 [hep-th/0008001].}
%%CITATION = HEP-TH 0007018;%%
%%CITATION = HEP-TH 0008001;%%

\lref\kleb{I.~R.~Klebanov and N.~A.~Nekrasov, ``Gravity duals of fractional
branes and logarithmic RG flow'', Nucl.\ Phys.\ B {\bf 574} (2000) 263
[hep-th/9911096]. I.~R.~Klebanov and A.~A.~Tseytlin, ``Gravity duals of
supersymmetric SU(N) x SU(N+M) gauge theories'', Nucl.\ Phys.\ B {\bf 578}
(2000) 123 [hep-th/0002159]. ~K.~Oh and R.~Tatar, ``Renormalization group
flows on D3 branes at an orbifolded conifold'', JHEP{\bf 0005} (2000) 030
[hep-th/0003183]. ~I.~R.~Klebanov and M.~J.~Strassler,
``Supergravity and a confining gauge theory: Duality cascades and
(chi)SB-resolution of naked singularities'', JHEP{\bf 0008} (2000) 052
[hep-th/0007191]. ~C.~P.~Herzog and I.~R.~Klebanov, ``Gravity duals of
fractional branes in various dimensions'', hep-th/0101020.}
%%CITATION = HEP-TH 9911096;%%
%%CITATION = HEP-TH 0002159;%%
%%CITATION = HEP-TH 0003183;%%
%%CITATION = HEP-TH 0007191;%%
%%CITATION = HEP-TH 0101020;%%

\lref\bst{H.~J.~Boonstra, K.~Skenderis and P.~K.~Townsend, ``The domain
wall/QFT correspondence'', JHEP{\bf 9901} (1999) 003 [hep-th/9807137].}
%%CITATION = HEP-TH 9807137;%%

\lref\cvetic{M.~Cvetic, G.~W.~Gibbons, H.~Lu and C.~N.~Pope, ``Ricci-flat
metrics, harmonic forms and brane resolutions'', hep-th/0012011. ~M.~Cvetic,
G.~W.~Gibbons, H.~Lu and C.~N.~Pope, ``Supersymmetric non-singular
fractional D2-branes and NS-NS 2-branes'', hep-th/0101096.}
%%CITATION = HEP-TH 0012011;%%
%%CITATION = HEP-TH 0101096;%%

\lref\bvs{M.~Bershadsky, C.~Vafa and V.~Sadov, ``D-Branes and Topological
Field Theories'', Nucl.\ Phys.\ B {\bf 463} (1996) 420 [hep-th/9511222].}
%%CITATION = HEP-TH 9511222;%%

\lref\bjsv{M.~Bershadsky, A.~Johansen, V.~Sadov and C.~Vafa, ``Topological
reduction of 4-d SYM to 2-d sigma models'', Nucl.\ Phys.\ B {\bf 448} (1995)
166 [hep-th/9501096].}
%%CITATION = HEP-TH 9501096;%%

\lref\fs{A.~Fayyazuddin and D.~J.~Smith, ``Warped AdS near-horizon geometry
of completely localized intersections of M5-branes'', JHEP {\bf 0010}
(2000) 023 [hep-th/0006060].}
%%CITATION = HEP-TH 0006060;%%

\lref\bfms{B.~Brinne, A.~Fayyazuddin, S.~Mukhopadhyay and D.~J.~Smith,
``Supergravity M5-branes wrapped on Riemann surfaces and their QFT duals'',
JHEP {\bf 0012} (2000) 013 [hep-th/0009047]. ~B.~Brinne, A.~Fayyazuddin,
T.~Z.~Husain and D.~J.~Smith, ``N = 1 M5-brane geometries'', hep-th/0012194.}
%%CITATION = HEP-TH 0012194;%%
%%CITATION = HEP-TH 0009047;%%

\lref\ens{See, for example, J.~D.~Edelstein, C.~N\'u\~nez and F.~A.~Schaposnik,
``Bogomol'nyi Bounds and Killing Spinors in d=3 Supergravity'', Phys.\
Lett.\ B {\bf 375} (1996) 163 [hep-th/9512117].}
%%CITATION = HEP-TH 9512117;%%

\lref\imsy{N.~Itzhaki, J.~M.~Maldacena, J.~Sonnenschein and S.~Yankielowicz,
``Supergravity and the large N limit of theories with sixteen supercharges'',
Phys.\ Rev.\ D {\bf 58} (1998) 046004 [hep-th/9802042].}
%%CITATION = HEP-TH 9802042;%%

\lref\vafa{C.~Vafa, ``Superstrings and topological strings at large N'',
hep-th/0008142.}
%%CITATION = HEP-TH 0008142;%%

\lref\bp{H.~Partouche and B.~Pioline, ``Rolling among G(2) vacua,''
JHEP{\bf 0103} (2001) 005 [hep-th/0011130].}
%%CITATION = HEP-TH 0011130;%%

\lref\civ{F.~Cachazo, K.~Intriligator and C.~Vafa, ``A Large N Duality via
a Geometric Transition'', hep-th/0103067.}
%%CITATION = HEP-TH 0103067;%%

\lref\bsa{B.~S.~Acharya, ``On realising N = 1 super Yang-Mills in M
theory'', hep-th/0011089.}
%%CITATION = HEP-TH 0011089;%%

\lref\ach{B.~S.~Acharya, ``Confining strings from G(2)-holonomy
spacetimes'', hep-th/0101206.}
%%CITATION = HEP-TH 0101206;%%

\lref\av{B.~Acharya and C.~Vafa, ``On domain walls of N = 1 supersymmetric
Yang-Mills in four dimensions'', hep-th/0103011.}
%%CITATION = HEP-TH 0103011;%%

\lref\gomis{J.~Gomis, ``D-Branes, Holonomy and M-Theory'', hep-th/0103115.}
%%CITATION = HEP-TH 0103115;%%

\lref\cdlo{P.~Candelas and X.~C.~de la Ossa, ``Comments On Conifolds'',
Nucl.\ Phys.\ B {\bf 342} (1990) 246.}
%%CITATION = NUPHA,B342,246;%%

\lref\pzt{L.~A.~Pando Zayas and A.~A.~Tseytlin, ``3-branes on resolved
conifold'', JHEP{\bf 0011} (2000) 028 [hep-th/0010088].}
%%CITATION = HEP-TH 0010088;%%

\lref\bs{R.~Bryant and S.~Salamon, ``On the construction of some complete
metrics with exceptional holonomy'', Duke\ Math.\ J.\ {\bf 58} (1989) 829.}

\lref\gpp{G.~W.~Gibbons, D.~N.~Page and C.~N.~Pope, ``Einstein Metrics On
$S^3$, $R^3$ And $R^4$ Bundles'', Commun.\ Math.\ Phys.\ {\bf 127} (1990)
529.}
%%CITATION = CMPHA,127,529;%%

\lref\aw{M.~Atiyah and E.~Witten, to appear. E.~Witten, talk given at {\it
Heterotic Dreams and Asymptotic Visions: The 60th Birthday Celebration for
David Gross}, ITP, Santa Barbara, March 2--3, 2001.}

%%%%%%%%%%
\Title{\vbox{\baselineskip12pt
\hbox{HUTP-01/A014}
\hbox{hep-th/0103167}
}}
{\vbox{\centerline{D6 branes and M--theory geometrical transitions}
\medskip
\centerline{from gauged supergravity}}}
\centerline{Jos\'e D. Edelstein and Carlos N\'u\~nez}

\bigskip
\medskip
{\vbox{\centerline{\sl Lyman Laboratory of Physics, Harvard University}
\centerline{\sl Cambridge, MA 02138, USA}
\centerline{\it edels,nunez@lorentz.harvard.edu}}}

\bigskip
\bigskip
\noindent
We study the supergravity duals of supersymmetric theories arising in the
world--volume of D6 branes wrapping holomorphic two--cycles and special
Lagrangian three--cycles within the framework of eight dimensional gauged
supergravity. When uplifted to 11d, our solutions represent M--theory on
the background of, respectively, the small resolution of the conifold and
a manifold with $G_2$ holonomy. We further discuss on the flop and other
possible geometrical transitions and its implications.

\Date{20 March 2001}

%\listtoc \writetoc

%%%%%%%%%%%%%%%%%%%%%%%%%%%%%%%%%%%%%%%
%%%%%%%%%%
%%%%%%%%%%   Section 1
%%%%%%%%%%
%%%%%%%%%%%%%%%%%%%%%%%%%%%%%%%%%%%%%%%
\newsec{Introduction}

The world--volume low--energy dynamics of D--branes in certain curved
backgrounds defines a topologically twisted supersymmetric field theory \bvs.
The twisting is necessary to allow for the world--volume of the brane to
support covariantly constant spinors (this is reminiscent of a similar
phenomenon in lower dimensional supergravities \ens). If the D--brane is
wrapping a nontrivial cycle, and we take its size to zero, the infrared
dynamics of the system is described by a lower dimensional field theory 
with either {\it ordinary} or {\it twisted} (depending on the higher
dimensional twisting being respectively {\it partial} or {\it full})
{\it reduced} supersymmetry \bjsv. The amount of supersymmetry preserved
has to do with the way in which the cycle is embedded in a higher
dimensional space. If the number of branes is taken to be large, this
sort of systems provide a supergravity dual description of $\CN = 1$ or
$\CN = 2$ supersymmetric field theories \mnu\kleb\agk\bfms\cvetic.

In this paper we will consider D6 brane configurations that reduce, at low
energies, to theories with four and eight supercharges in four and five
dimensions. The D6 brane system is best described in the infrared by
means of $\CN = 2$ seven dimensional super Yang--Mills theory \imsy. So,
for example, wrapping these branes on $S^3$ would imply, after appropriate
twisting, breaking one quarter of the supercharges, the theory reducing
to pure $\CN = 1$ four dimensional super Yang--Mills in the infrared. The
above referred twisting corresponds to $S^3$ being a special Lagrangian
submanifold of a Calabi--Yau threefold, namely, the deformed conifold
$T^*S^3$. On the other hand, if the D6 branes wrap a holomorphic $S^2$ in
the cotangent bundle of $S^2$, $T^*S^2$, the infrared dynamics will be
governed by five dimensional $\CN = 2$ super Yang--Mills theory.

It was recently proposed that the configuration of D6 branes wrapping an
$S^3$ in $T^*S^3$ is dual, through a conifold transition, to a type IIA 
geometry where the D6 branes have dissapeared, being replaced by RR
fluxes on the blownup $S^2$ \vafa.
Conversely, there is a mirrored type IIB version of this phenomenon with D5
branes wrapping the $S^2$ becoming RR fluxes on the $S^3$. It was almost
immediately realized that this duality can be better viewed in M--theory on
$G_2$ holonomy manifolds \bsa, where it corresponds to a flop transition \amv.
(See also \bp\ for a recent discussion on topology change in $G_2$ holonomy
manifolds.) It is natural to analyze these configurations in 11d for the
fact that uplifted D6 branes become purely gravitational. Besides, the
D6 branes are strongly coupled in the ultraviolet and the would be
decoupling limit has to be addressed in eleven dimensions. In particular,
the 11d supergravity solution is trustable for any number of branes.
Another difference with other D--branes is given by the fact that massive
geodesics can escape to infinity signaling the non decoupling of gravity
\imsy.

It is our purpose in this paper to study this sort of solutions under the
light of lower dimensional gauged supergravity. This is the natural
framework to perform twisting. The solutions emerging from these theories
correspond to near horizon D--brane solutions thus giving directly the
gravity duals of gauge theories living on the world--volume of the brane. 
Since we will work with D6 branes, the twisting would require to impose
boundary conditions on eight dimensional gravitational, gauge and scalar
fields so, following the methods introduced in \mnu, the natural set up for
this problem is eight dimensional gauged supergravity. In particular, we
will work within the framework of {\it maximal} $8d$ gauged supergravity
\ss\ so as to have enough room for different twistings. The virtue of
gauged supergravities in this respect is that they provide quite cleanly
the gauge field modes that undertake the partial twisting.

Uplifting to eleven dimensions will leave us with $M$--theory on Ricci
flat backgrounds corresponding to the small resolution of the conifold and
a $G_2$ holonomy manifold. Both manifolds eventually develop singularities
where transitions to a different manifold might be possible. In the latter
case, for example, the geometrical transition correspond to the above referred
flop between two three--spheres that, at the singular point, constitute
the base of a cone \amv. Instead, in the former case, we found that there
is no transition, and the theory in the ultraviolet falls into the
singularity. The reason for the absence of a geometrical transition can be
attributed, as we will discuss, to the non existence of a $\theta$ angle in
five dimensional theories. This suggest that a duality between large $N$ five
dimensional $\CN = 2$ super Yang--Mills theory and superstrings propagating
in a K3 manifold with fluxes turned on, in the spirit of \vafa, does not
take place.

The plan of the paper is as follows. In Section 2 we review maximal gauged
supergravity in eight dimensions and prepare the set up for the search of
solutions. Section 3 is devoted to the case of D6 branes wrapping special
Lagrangian $3$--cycles. We first construct solutions of 8d gauged
supergravity that are subsequently uplifted to 11d. The resulting geometry
is that of a $G_2$ holonomy manifold recently studied in \amv. In section
4 we consider the case of D6 branes on holomorphic $2$--cycles in a
deformed $A_1$ singularity of $K3$, namely $T^*S^2$. When uplifted to 11d
our solution is the small resolution of the conifold $\CO(-1) + \CO(-1) \to
\IP^1$. We discuss on the obstructions to the geometrical transitions
appearing in this case and their relation to generic aspects of five
dimensional gauge theories. We conclude in section 5 with a discussion of
our results, and an outlook of avenues for further research.
\medskip

{\it Note Added:} While the final version of this paper was being
typewritten, some results that overlap part of ours were reported by
Jaume Gomis \gomis.

%%%%%%%%%%%%%%%%%%%%%%%%%%%%%%%%%%%%%%%
%%%%%%%%%%%
%%%%%%%%%%% Section 2
%%%%%%%%%%%
%%%%%%%%%%%%%%%%%%%%%%%%%%%%%%%%%%%%%%%
\newsec{Review of $d=8$ gauged supergravity}

Maximal gauged supergravity in eight dimensions was originally constructed
by Salam and Sezgin \ss. It arises from dimensional reduction of $11d$
supergravity on a $SU(2)$ group manifold \scs. The field content of this
theory consists of the metric $g_{\mu\nu}$, a dilatonic scalar $\Phi$, five
scalars given by a unimodular $3 \times 3$ matrix $L_\a^i$ in the coset
$SL(3,\IR)/SO(3)$, a seventh scalar $B$, a three--form $B_{(3)}$, three
two--forms $B^i_{(2)}$, three vector fields $B^i_{(1)}$ and a $SU(2)$ gauge
potential $A^i_\mu$, as well as the pseudo Majorana spinors $\psi_\mu$ and
$\chi_i$. In this paper we are going to restrict ourselves to a sector of
the theory with vanishing $B$--fields. This amounts to pure gravitational
solutions of the $11d$ system. The bosonic dynamics in this sector is
governed by the Lagrangian
\eqn\boslag{
e^{-1} \CL = {1 \over 4} R - {1 \over 4} e^{2\Phi} F_{\mu\nu}^{~i}
F^{\mu\nu i} - {1 \over 4} P_{\mu ij} P^{\mu ij} - \half (\p_\mu\Phi)^2
- {g^2 \over 16} e^{-2\Phi} (T_{ij} T^{ij} - \half T^2) ~,}
where $e$ is the determinant of the {\it achtbein} $e_\mu^a$, $F_{\mu\nu}^i$
is the Yang--Mills field strength and $P_{\mu ij}$ is a symmetric and
traceless quantity defined by
\eqn\pyq{
P_{\mu ij} + Q_{\mu ij} \equiv L_i^\a (\p_\mu \d_\a^{~\b} - g \e_{\a\b\g}
A_\mu^\g) L_{\b j} ~,}
$Q_{\mu ij}$ being the antisymmetric counterpart. We have set $\kappa = 1$.
As usual, greek indices are curved ($\a,\b,\dots$ are in the group manifold
\foot{While working in eight dimensions, these indices describe a flat
space. The dependence on the coordinates of the group manifold have been
factored out, and it will only reappear when uplifting to $11d$ is
performed.} and $\mu,\nu,\dots$ label space--time coordinates) while latin
ones are flat. Notice that, for example, $A_\mu^\g = L_i^\g A_\mu^i$, as
well as $F_{\mu\nu}^{~\a} = L_i^\a F_{\mu\nu}^i$. Finally, the potential
energy corresponding to the scalar fields is governed by the so-called
$T$--tensor,
\eqn\tij{
T^{ij} = L_\a^i L_\b^j \d^{\a\b} ~,}
and $T = \d_{ij} T^{ij}$. The equations of motion are
\eqn\eins{
R_{\mu\nu} = P_{\mu ij} P_\nu^{ij} + 2 \p_\mu\Phi \p_\nu\Phi + 2 e^{2\Phi}
F_{\mu\g}^i F_\nu^{~\g i} - {1 \over 3} g_{\mu\nu} \nabla^2 \Phi ~,}
\eqn\maxw{
\nabla_\mu (e^{2\Phi} F^{\mu\nu i}) = - e^{2\Phi} P_\mu^{ij} F_{~j}^{\mu\nu}
- g g^{\nu\g} \e^{ijk} P_{\g jl} T_k^{~l} ~,}
\eqn\epij{
\nabla_\mu P^{\mu ij} = - {2 \over 3} \delta^{ij} \nabla^2 \Phi + e^{2\Phi}
F_{\mu\nu}^i F^{\mu\nu j} + {g^2 \over 2} e^{-2\Phi} \Theta^{ij} ~,}
where $\Theta^{ij}$ is short for
\eqn\thij{
\Theta^{ij} \equiv T^i_{~k} T^{jk} - \half T T^{ij} - \half \d^{ij} 
\bigl( T_{kl} T^{kl} - \half T^2 \bigr) ~.}
Notice that the dilaton equation is obtained from \epij\ by tracing over
the latin indices.

The supersymmetry transformations for the fermions are given by
\eqn\stpsi{
\d\psi_\g = \CD_\g \e + {1 \over 24} e^{\Phi} F_{\mu\nu}^i \G_i
(\G_\g^{~\mu\nu} - 10 \d_\g^{~\mu} \G^\nu) \e - {g \over 288} e^{-\Phi}
\e_{ijk} \G^{ijk} \G_\g T \e ~,}
\eqn\stchi{
\d\chi_i = \half (P_{\mu ij} + {2 \over 3} \d_{ij} \p_\mu\Phi) \G^j
\G^\mu \e - {1 \over 4} e^{\Phi} F_{\mu\nu i} \G^{\mu\nu} \e - {g \over 8}
e^{-\Phi} (T_{ij} - \half \d_{ij} T) \e^{jkl} \G_{kl} \e ~,}
where the covariant derivative is
\eqn\covd{
\CD_\mu \e = \bigl( \p_\mu + {1 \over 4} \o_\mu^{ab} \G_{ab} + {1 \over 4}
Q_{\mu ij} \G^{ij} \bigr) \e ~.}
It is useful for later purposes to work alternatively with spinors of
$32$ components or doublets of sixteen components. Then, we will use the
following representation for the Clifford algebra
\eqn\cliff{
\G^a = \g^a \times \II ~~~~~~~ \G^i = \g_9 \times \s^i ~,}
where $\g^a$ are eight dimensional gamma matrices ($a$ being a flat index),
$\g_9 = i \g^0 \g^1 \dots \g^7$, with $\g_9^2 = 1$, and $\s^i$ are the
Pauli matrices corresponding to the $R$--symmetry group. It will be
finally convenient to introduce $\G_9 \equiv {1 \over 6i} \e_{ijk} \G^{ijk}
= \g_9 \times \II$.

In the following we will consider supergravity duals of gauge theories in
four and five dimensions with four and eight supercharges respectively. Our
procedure is based on taking the low energy limit for a D6 brane wrapped on
three and two supersymmetric cycles in Calabi--Yau and $K3$ manifolds. The
structure group of the normal bundle of these cycles is, respectively,
$SO(3)$ and $SO(2)$, thus Salam--Sezgin theory has enough room for their
twisting. When the energies are low enough, the cycle decouples and we
remain with a theory that has less dimensions and less supersymmetries
than the original one.

Since we will work with  D6 branes, it seems natural to consider seven
dimensional boundary conditions for gauge and scalar fields, so, the
natural set up for this problem is eight dimensional gauged supergravity.
We can see that the vacuum supersymmetric solution of this theory is given
by \bst\
\eqn\metr{
ds_8^2 = e^{{2 \over 3} \phi} dx^2_{1,6} + dr^2 ~,}
\eqn\dila{
e^{\phi-\phi_0} = r ~,}
where $\phi_0 = \log ({3g \over 8})$. When uplifted to eleven dimensions
by means of the prescription given in Ref.\ss, after appropriate coordinate
rescaling, the higher dimensional configuration is
\eqn\solucion{
ds^2 = dx^2_{1,6} + N (d\rho^2 + \rho^2 d\O_3) ~.}
After modding out the outer three--sphere by $\IZ_N$, we get an ALE space
with an $A_{N-1}$ singularity in coincidence with the uplifting of the near
horizon solution corresponding to D6 branes in type IIA \imsy.

%%%%%%%%%%%%%%%%%%%%%%%%%%%%%%%%%%%%%%
%%%%%%%%%%%
%%%%%%%%%%%  Section 3
%%%%%%%%%%%
%%%%%%%%%%%%%%%%%%%%%%%%%%%%%%%%%%%%%%
\newsec{D6 branes on the deformed conifold}

In this section we will obtain the gravity dual of $\CN = 1$ super
Yang--Mills theory in four dimensions, arising in the low--energy dynamics
of D6 branes wrapped on $S^3$ in $T^*S^3$, starting from eight dimensional
gauged supergravity. Let us start with an ansatz for the metric that
describes such deformation of the world--volume of the D6 brane
\eqn\metrica{
ds^2 = e^{2f} dx^2_{1,3}+ e^{2h} d\O_3 + dr^2 ~,}
where  $d\O_3$ is the metric of the unit three--sphere. As explained in the
introduction, wrapping the branes on a curved cycle implies that the theory
has to be twisted on the curved part.

The fields on the D6 branes transform under $SO(1,6) \times SO(3)_R$ as
({\bf 8},{\bf 2}) for the fermions and ({\bf 1},{\bf 3}) for the scalars,
while the gauge field is a singlet under $R$--symmetry. When we wrap the
D6 branes on a three--cycle, the symmetry group splits as $SO(1,3) \times
SO(3) \times SO(3)_R$, and we shall construct a diagonal $SO(3)_D$ from
the $SO(3)$ of the cycle and the one of the R-symmetry (in other words,
we mix the spin conection with the gauge connection, as explained above).
It can be easily seen that the effect of the twisting is to preserve the
vector fields but transforms the scalars in one forms on the curved
surface, so we are left with a theory with no scalars fields in the
infrared; besides four supercharges are preserved. 

We will describe the $S^3$ as a $SU(2)$ group manifold by means of the left
invariant forms $w^i$,
\eqn\forms{\eqalign{
w^1 = & \cos\ph ~d\t + \sin\t \sin\ph ~d\psi ~, \cr
w^2 = & \sin\ph ~d\t - \sin\t \cos\ph ~d\psi ~, \cr
w^3 = & d\ph + \cos\t ~d\psi ~,}}
satisfying
\eqn\relfor{
dw^i = \half \e^{ijk} w^j \wedge w^k ~,}
in terms of which the metric of the unit sphere simply reads
\eqn\unitsph{
d\O_3 = {1 \over 4} \sum_{i=1}^3 (w^i)^2 ~.}
The twisting is achieved by turning on a non--Abelian $SO(3)$ gauge field
given by the left invariant form of the three sphere,
\eqn\amu{
A^i = - (2g)^{-1} w^i ~,}
whose field strength
\eqn\fmunu{
F^i = - (8g)^{-1} \e^{ijk} w^j \wedge w^k ~,}
trivially obeys the corresponding equation of motion. This correponds to a
complete identification of the spin connection with the R--symmetry. In
such case it is possible to get rid of the scalars $L_\a^i$,
\eqn\scalu{
L_\a^i = \d_\a^i ~~~ \Rightarrow ~~~ P_{ij} = 0 ~~, ~~Q_{ij} = - g \e_{ijk}
A^k ~.}
The $T$--tensor is drastically simplified to $T_{ij} = \d_{ij}$, $T = 3$,
and $\Theta_{ij} = {1 \over 4} \d_{ij}$. Supersymmetric configurations
require the following projections in the parameter $\e$:
\eqn\proju{
\g_\ur ~\e = -i \g_9 ~\e ~~~~~~~ \g_{ab} ~\e = - \s^{ab} \e ~,}
where $a,b = \ut, \up, \ups \equiv 1, 2, 3$~ are the directions along the
three--sphere. These projection leave unbroken $1/8$ of the original
supersymmetries, that is, four supercharges. The first order BPS equations
are,
\eqn\buu{
f' = {1\over 3}\Phi' = - {1 \over 2g} e^{\Phi - 2h} + {g \over 8}
e^{-\Phi} ~,}
\eqn\bud{
h' = {3 \over 2g} e^{\Phi - 2h} + {g \over 8} e^{-\Phi} ~.}
By simple inspection, we can quickly find a solution
\eqn\solu{
e^{2 \Phi} = {g^2 \over 16} r^2 ~, ~~~~~
e^{2 h} = {3 \over 4} r^2 ~.}
Notice that the relation $\Phi' = 3 f'$ is forced from the Ricci flatness
of the corresponding eleven dimensional solution. When uplifted to 11d, we
obtain a Ricci flat solution of the form $M_4 \times Y_7$ where $Y_7$ is a
cone whose base $X_6$ is an Einstein manifold with the topology of $S^3
\times \tS^3$,
\eqn\soluu{
ds^2 = dx^2_{1,3} + dr^2 + {r^2 \over 9} \bigl[(w^a)^2 + (\tw^a)^2
- w^a \tw^a \bigr] ~,}
$\tw^a$ being the left invariant one forms associated with $\tS^3$.
This metric coincides with the asymptotic at large $r$ of the $M$--theory
solution on a $G_2$ holonomy manifold studied in Ref.\amv, and we note
that the solution is singular in the infrared. It is natural to try to
obtain a solution where the singularity is absent. In this system we do
not have further degrees of freedom to turn on, that could occasionally
solve the singularity; this  means that there must be other solutions to
the BPS equations \buu\bud, such that, when uplifted to eleven dimensions,
do not give place to singularities in the infrared.

We can define indeed a pair of functions, $u \equiv h + \Phi$ and $v
\equiv 3 h - \Phi$, the system simplifies to $e^u du = {g^2 \over 12} e^v
dv$, whose immediate solution is
\eqn\soluyv{
e^u = {g^2 \over 12} \Bigl( e^v - {a^3 \over 3^{3 \over 2}} \Bigr) ~,}
$a$ being a constant. There is an amusing change of variable
\eqn\changu{
r(\rho) = {(2 g)^{1 \over 2} \over 3^{3 \over 2}} \Bigl( \rho^{3 \over 2}
~_2\! F_1[-{1 \over 2};{1 \over 4},{1 \over 2};{a^3 \over \rho^3}] - a^{3
\over 2} {\sqrt{\pi} \G ({3 \over 4}) \over \G ({1 \over 4})} \Bigr) ~,}
where $_2\! F_1[a,b,c,z]$ is the hypergeometric function \foot{Notice that
in our case, for $\rho \geq a$, it has a real variable $z \leq 1$ such that
the change of variables has not branch cut discontinuity. The substracted
constant in \changu\ just amounts to $r(a)=0$.}
\eqn\hyper{
_2\! F_1[a,b,c,z] = \sum_{m=0}^{\infty} {(a)_{m} (b)_m \over (c)_{m}} ~{z^m
\over m!} ~,}
with $(a)_n = \G(a+n)/\G(a)$ the Pochhammer symbol, that allows us to find
a solution to the BPS equations of the form
\eqn\solubps{
e^{2 \Phi} = {g^3 \over 216} \rho^3 \Bigl( 1 - {a^3 \over \rho^3}
\Bigr)^{3
\over 2} ~~~
e^{2 h} = {g \over 18} \rho^3 \Bigl( 1 - {a^3 \over \rho^3} \Bigr)^{{1
\over 2}}.}
When  uplifted to 11 dimensions, we obtain
\eqn\met{
ds^2 = dx^2_{1,3} + {1 \over \Bigl( 1- {a^3 \over \rho^3} \Bigr)} d\rho^2
+ {\rho^2 \over 12} (\tw^a)^2 + {\rho^2 \over 9} \Bigl( 1 - {a^3 \over
\rho^3} \Bigr) \Bigl[ w^a - {1 \over 2} \tw^a \Bigr]^2 ~.}
This is the metric of a $G_2$ holonomy manifold which is topologically
$\IR^4 \times S^3$, originally constructed in \bs\gpp. It turns out from
our results that it can be obtained by means of an uplifting to 11d of a
solution of eight dimensional gauged supergravity. This fits nicely with
the discussions in \amv, in the sense that one expects gauged supergravity
to give a good description of the near horizon dynamics of D--branes, and
this is what we are presently making manifest.

The solution obtained so far represents an $M$--theory background which is 
the direct product of four dimensional Minkowski space and a $G_2$ holonomy
manifold. In order to understand the physics behind this configuration we
can go to type IIA. This issue was discussed in detail recently by Atiyah,
Maldacena and Vafa \amv, and we briefly remind here the main aspects of
their discussion. The radial variable in \met\ $\rho \geq a$ fills $S^3$
while the other sphere $\tS^3$ remains of finite volume $a^3$ when the
former shrinks (see Fig.1). The $G_2$ holonomy manifold has isometry
group $SU(2)_L \times SU(2)_{\tilde L} \times SU(2)_D$, the first two
factors corresponding to the left action on $S^3$ and $\tilde S^3$
respectively, and the last one is the diagonal subgroup of $SU(2)_R
\times SU(2)_{\tilde R}$. There is a flop transition in which the two
spheres are exchanged. In this case, $M$--theory smooths out the
singularity thanks to the existence of $C$--field fluxes through the
three--sphere.

\bigskip
\centerline{\vbox{\hsize=5in\tenpoint
\psfig{figure=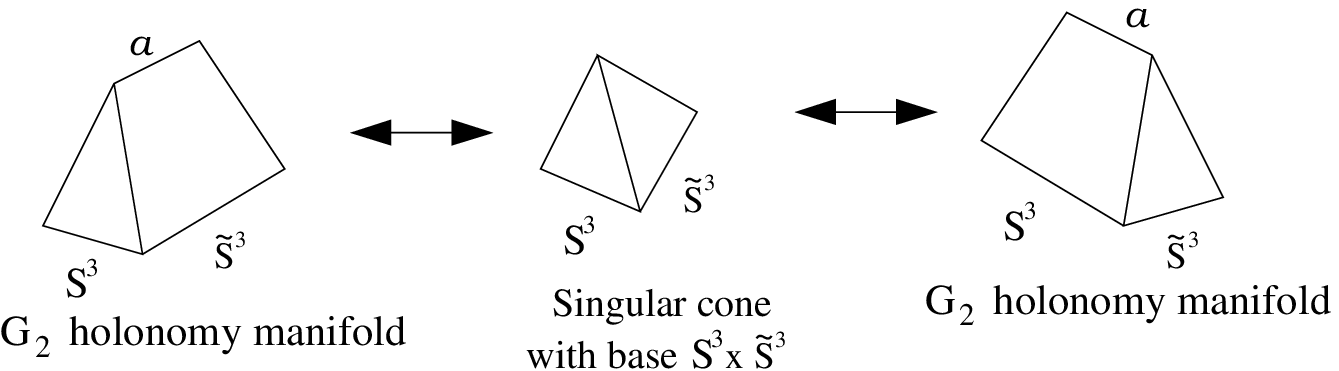}
\vglue.3in
Fig. 1. Flop transition in M--theory. The $G_2$ holonomy manifold on the
left can be deformed to the one on the right. $M$--theory avoids the
singular point due to the existence of $C$--field fluxes. }}
\vskip 3mm

There are two very different quotients of this manifold: a singular one
by modding out by $\IZ_N \subset U(1) \subset SU(2)_L$, and a
non--singular quotient if one instead chooses $\IZ_N \subset U(1) \subset
SU(2)_{\tilde L}$. This is due to the fact that $S^3$ shrinks to a point
when $\rho \to a$ while $\tilde S^3$ has radius $a$. Modding
out by $\IZ_N \subset U(1) \subset SU(2)_L$ results in an $A_{N-1}$
singularity fibered over $\tS^3$ so, after KK reduction along the circle
corresponding to the $U(1)$, one ends with $N$ D6 branes wrapped on $\tS^3$
whose normal bundle is that corresponding to the three--sphere being a
special Lagrangian in the deformed conifold \amv. The holonomy gets
correspondingly reduced from $G_2$ to $SU(3)$ \gomis. The second case,
amounts to modding out by $\IZ_N \subset U(1) \subset SU(2)_\tL$, which
has no fixed points so the quotient is regular. After KK reduction one ends
with a non--singular type IIA configurations (without D--branes) on a space
with the topology of $\CO(-1) + \CO(-1) \to \IP^1$ \amv, and with $N$ units
of RR flux through the finite radius $S^2$. This allowed the authors of
Ref.\amv\ to obtain the duality proposed by Vafa \vafa\ from a geometrical
flop in 11d.

The gauge theory defined in the planar directions of the D6 branes is, as
we pointed out above, a four dimensional $\CN = 1$ super Yang--Mills theory
(possibly with extra light fields corresponding to KK modes). We will
analize the gauge theory from a IIA perspective. The coupling constant
of this gauge theory is given by
\eqn\coupling{
{1 \over g_{4}^2} = {{\rm Vol}\tS^3 \over g_{7}^2} ~.}
Following \vafa, we can define in string theory a coupling superfield whose
lower component is
\eqn\coupfield{
Y = {V \over g_s} + i C ~,}
where $V$ represents the volume of the three sphere in the string frame
and $C$ is the value of the three form potential of IIA that plays the
role of the $\theta$ angle in the gauge theory. Besides, we introduce the
usual gauge superfield $S = g_s Tr W^2$, so that the action of the theory
will be given by
\eqn\action{
\CS = {1 \over g_s} \int d^2\theta ~Y S ~.}
The instantons, that in our case will be euclidean D2 branes wrapped on
the three--sphere, will correct this low energy action
\eqn\actioncorrected{
\CS_{inst} \propto \int d^2\theta N^2 e^{Y/N} ~.}
After extremizing with respect to the superfield $Y$ and $S$, one obtains
a superpotential where it can be seen that the breaking of $U(1)$ to $Z_2$
has taken place, due to the appearence of $N$ vacua and a gaugino condensate.

%%%%%%%%%%%%%%%%%%%%%%%%%%%%%%%%%%%%%%%%%%%%
%%%%%%%%%%
%%%%%%%%%%    Section 4
%%%%%%%%%%
%%%%%%%%%%%%%%%%%%%%%%%%%%%%%%%%%%%%%%%%%%%%
\newsec{D6 branes on $S^2$ in $T^*S^2$}

In the same vein of previous section, we now wrap $N$ D6 branes on a
holomorphic $S^2$ in $T^*S^2$. The corresponding split of the symmetry
group is $SO(1,4) \times SO(2) \times SU(2)_R$ such that, after twisting
the $SO(2)$ with a $U(1) \subset SU(2)_R$, the preserved fermions are those
transforming in the $(4,\pm,\mp)$ representations, {\it i.e.} there are
eight supercharges. Notice that out of three scalars, the one corresponding
to the particular $U(1)$ chosen in the $R$--symmetry group survives the
twisting. This leaves us with the field content of $\CN = 2$ five
dimensional super Yang--Mills theory. The real scalar parametrizes the
Coulomb branch of the theory.

From the gauged supergravity point of view, the twisting of this
configuration allows us to get rid of all but one
scalar field $\la$ that enters $L_\a^i$ as
\eqn\scald{
L_\a^i = {\rm diag}(e^\la,e^\la,e^{-2\la}) ~,}
such that the symmetric $P_{ij}$ and antisymmetric $Q_{ij}$ are
\eqn\pandq{
P_{ij} = \pmatrix{
\p\la & 0 & -g \sinh 3\la ~ A^2 \cr
 & \p\la & g \sinh 3\la A^1 \cr 
 &  & -2 \p\la} ~~~~~
Q_{ij} = \pmatrix{
0 & -g A^3 & g \cosh 3\la A^2 \cr
 & 0 & -g \cosh 3\la A^1 \cr
 &  & 0} ~.}
It is enough to perform the twisting to turn on the gauge field $A^3$. The
ansatz for the metric is
\eqn\ansd{
ds^2 = e^{2f} dx^2_{1,4} + e^{2h} d\O_2 + dr^2 ~,}
where $d\O_2 = d\t^2 + \sin^2\t d\phi^2$, and $f$ and $h$ are only
functions of the radial coordinate. Imposition of the BPS conditions $\d_\e
\psi_\mu = \d_\e \chi_i = 0$ translates into the following projections of
the supersymmetric parameter
\eqn\projd{
\g_{\ut\up} ~\e = - \G^{12} ~\e ~~~~~~~ \g_{\ur} ~\e = - i\g_9 ~\e ~.}
On the other hand, the BPS equations are:
\eqn\bdu{
f' = {1\over 3}\Phi' = - {1 \over 6g} e^{\Phi - 2\la - 2h} + {g \over
24} (2 e^{2\la} + e^{-4\la}) e^{-\Phi} ~,}
\eqn\bdd{
h' = {5 \over 6g} e^{\Phi - 2\la - 2h} + {g \over 24} (2 e^{2\la} +
e^{-4\la}) e^{-\Phi} ~,}
\eqn\bdt{
\la' = {1 \over 3g} e^{\Phi - 2\la - 2h} - {g \over 6} (e^{2\la} -
e^{-4\la}) e^{-\Phi} ~.}
Let us first present a solution in the simplest case of constant $\la$. It
is clear from \bdt\ that $\la > 0$ and $e^\Phi = \xi^{1/2} e^h$ with $\xi =
{g^2 \over 2} (e^{4\la} - e^{-2\la})$. Plugging this relation back into 
\bdu\bdd\ it is immediate to obtain $\la = {1 \over 6} \log {3 \over 2}$
and $e^{2h} = {3 \over 8} r^2$. Uplifting to 11d leads, after a suitable
change of variables, to a metric of the form $M_5 \times Y_6$ where $Y_6$
is the singular conifold with base $T^{1,1}$,
\eqn\soldu{
ds^2 = dx^2_{1,4} + dr^2 + r^2 d\Sigma_{1,1} ~,}
\eqn\tuu{
d\Sigma_{1,1} = {1 \over 9} (d\psi + \sum_{a=1}^2 \cos\t_a d\phi_a)^2 + {1
\over 6} \sum_{a=1}^2 (d\t_a^2 + \sin^2\t_a d\phi_a^2) ~.} 
The fact that the solution in 11d involves a Calabi--Yau manifold is
expected not only from the Ricci flatness condition but for the fact that,
in 10d, the D6 branes are wrapping a holomorphic cycle of a $K3$ manifold
whose holonomy is $SU(2)$, and this uplifts to $SU(3)$ holonomy manifolds
\gomis. It is natural to attempt a resolution of the singularity. The small
resolution of the conifold has a metric given by \cdlo\pzt
\eqn\small{\eqalign{
ds^2_{\rm res} = & \kap(\rho)^{-1} d\rho^2 + {1 \over 9} \kap(\rho) \rho^2
(d\psi + \sum_{a=1}^2 \cos\t_a d\phi_a)^2 \cr
& + {1 \over 6} \rho^2 (d\t_1^2 + \sin^2\t_1 d\phi_1^2) + {1 \over 6} 
(\rho^2 + 6 a^2) (d\t_2^2 + \sin^2\t_2 d\phi_2^2) ~,}}
where
\eqn\kapp{
\kap(\rho) \equiv {\rho^2 + 9 a^2 \over \rho^2 + 6 a^2} ~.}
When $a \to 0$ the metric reduces to that of the standard conifold
\soldu\tuu. It is easy to see from \small\ that near the former apex of
the conifold the $S^3$
shrinks to zero while the $S^2$ remains of finite size $a^2$. To resolve
the singularity, we should excite new degrees of freedom. It is natural, in
this case, to look for excitations of the scalar field $\la$, that must be
the source of the resolution factor $\kap$. In the singular limit, as well
as for large radius, $\kap \to 1$, so it should enter as a logarithm, say,
\eqn\newla{
\la(r) = {1 \over 6} \Bigl( \log {3 \over 2} - \log\kap(r) \Bigr) ~.}
We now insert this function back in \bdu\bdd\bdt\ and peform a not less
amusing change of variables,
\eqn\chvd{
r(\rho) = {2^{17 \over 12} \over 3^{17 \over 12}} \rho^{3 \over 2} 
F_1[{3 \over 4};-{5 \over 12},{5 \over 12};{7 \over 4};-{\rho^2 \over 6
a^2},-{\rho^2 \over 9 a^2}] ~,}
where $F_1[a;b_1,b_2;c;x,y]$ is the Appell hypergeometric function of two
variables \foot{Notice that the variables of the Appel function are real
and negative so this change of variables is not singular neither it has
branch cut discontinuities.}
\eqn\appel{
F_1[a;b_1,b_2;c;x,y] = \sum_{m,n=0}^{\infty} {(a)_{m+n} (b_1)_m (b_2)_n
\over m! n! (c)_{m+n}} ~x^m y^n ~,}
finding a solution to the BPS equations of the form
\eqn\solbps{
e^{2 \Phi} = {g^3 \over 144} \rho^3 \kap^{1 \over 2}(\rho) ~~~
e^{2 h} = {3^{1 \over 3} g \over 2^{1 \over 3} 36} \rho (\rho^2 + 6 a^2)
\kap^{1 \over 6}(\rho) ~.}
Uplifting of this solution to 11d gives precisely the metric \small\ of the
small resolution of the conifold $\CO(-1) + \CO(-1) \to \IP^1$. It is again
remarkable that this solution was extracted cleanly from eight dimensional
gauged supergravity. In this case, the radial coordinate $\rho \geq 0$
fills the three--sphere while the two--sphere remains of finite radius when
$\rho \to 0$. The ALE $A_{N-1}$ singularity fibered over $S^2$, that
corresponds to $N$ D6 branes (after KK reduction along the appropriate
$U(1)$) in type IIA wrapping an $S^2$ in $T^*S^2$, is obtained after
modding out by $\IZ_N \subset \CO(-1) + \CO(-1)$. 

It is natural to ask at this point about the possibility of geometrical
transitions of the $\CO(-1) + \CO(-1) \to \IP^1$ manifold taking place.
There are some key differences with respect to the configuration of the
previous section. Indeed, the moduli space is one (real) dimensional,
being basically parametrized by the size of the sphere, that is, $a$. 
Thus, there is no possible conifold transition. Besides, in the present
case, there is no way to avoid the singular point $a = 0$. This is
related to the fact that these theories do not possess a $\theta$ term.
This further prevents geometrical transitions of the sort discussed in
section 2 of
\amv\ (in the context of type IIA). Moreover, when the system heads the
singularity, the gauge theory is flowing towards the ultraviolet. There
is no decoupling between the gauge theory and the string modes.

%%%%%%%%%%%%%%%%%%%%%%%%%%%%%%%%%%%%%%%%%%%%
%%%%%%%%%%
%%%%%%%%%%    Section 5
%%%%%%%%%%
%%%%%%%%%%%%%%%%%%%%%%%%%%%%%%%%%%%%%%%%%%%%
\newsec{Conclusions}

In the present paper we have obtained supergravity duals of D6 branes
wrapping holomorphic two--cycles in local $T^*S^2$ and special Lagrangian
submanifolds in a deformed conifold. More concretely, we focused on the
$M$--theory description of such configurations looking for geometric
transitions that might be pointing in the direction of new superstring
dualities that amount to large $N$ dualities of supersymmetric gauge theories.

It is interesting to remark that both metrics found in our paper, that is,
the small resolution of the conifold and the $G_2$ holonomy manifold, are
obtained from eight dimensional supergravity, which is in some sense the 
{\it natural} theory for the near horizon dynamics of the D6 branes. Notice
that our solutions correspond to the infinite coupling limit of type IIA,
due to the fact that our 10d dilaton diverges for large values of the
radial coordinates. Definitively, it would be of great interest to seek
for solutions where the string coupling goes to a constant.

Manifolds with $G_2$ holonomy give the appropriate M--theoretic background
corresponding to $\CN = 1$ four dimensional theories. It is then of great
interest to extend our results to include the other manifolds of this kind
that were reported in the literature \bs\gpp. From the string theory point
of view, it has been recently argued that they correspond to the uplifting
of space--time filling intersections of D6 branes \aw.

The framework provided by this sort of dualities that emerge from
geometrical transitions in $M$--theory seems very promising. In the last
few months, interesting papers addressing issues such as confining string
\ach, domain walls \av, gluino condensation and remarkable dualities between
different $\CN = 1$ gauge theories \civ\ have appeared. The information
provided by supergravity duals of these configurations is, in some sense,
complementary to the above mentioned dualities: unlike the latters, the
formers carry information of $D$ terms.

We think that gauged supergravities provide a natural framework to pose
some of these questions. Clearly, further research in the subject is
deserved.

%%%%%%%%%%%%%%%%%%%%%%%%%%%%%%%%%%%%%%%%%%%%%%%%
\medskip
\bigskip
\noindent
{\bf Acknowledgements:} 

We are very pleased to thank Jerome Gauntlett, Juan Maldacena, Kyungho Oh
and, especially, Cumrun Vafa for most valuable comments and discussions.
We thank Edward Witten for correspondence. The work of J.D.E. has been
supported by the Argentinian National Research Council (CONICET) and by a
Fundaci\'on Antorchas grant under project number 13671/1-55. The work of
C.N. has been supported by Fundaci\'on Antorchas.

\listrefs

\bye

\listrefs

\end